\documentclass[%
aip,
jcp,%
amsmath,amssymb,
reprint%
,longbibliography,
]{revtex4-1}

\usepackage{amsmath}
\usepackage{amsfonts}
\usepackage{amssymb,bm}
\usepackage{graphicx}
\usepackage{physics}
\usepackage{upgreek}
\usepackage[outercaption]{sidecap}   
\usepackage{xcolor}
\usepackage{hyperref}
\hypersetup{colorlinks=true,
	linkcolor=magenta,
	allcolors=magenta}
	\usepackage{newtxtext}
	\usepackage{newtxmath}
	\usepackage[version=4]{mhchem}
	\usepackage{blkarray}

	\DeclareMathAlphabet{\pazocal}{OMS}{zplm}{m}{n}

	\newcommand{\pL}{\ensuremath{\pazocal{L}}}
	\newcommand{\pK}{\ensuremath{\pazocal{K}}}

	\newcommand{\sing}{\ensuremath{\mathrm{S}}}
	\newcommand{\trip}{\ensuremath{\mathrm{T}}}

	\renewcommand{\op}[1]{\ensuremath{\hat{#1}}}

\begin{document}
		
		\title{Spin selective charge recombination in chiral donor-bridge-acceptor triads}
		\author{Thomas P. Fay\looseness=-1}
		\email{tom.patrick.fay@gmail.com}
		\affiliation{Department of Chemistry, University of California, Berkeley, CA 94720, USA\looseness=-1}
		\author{David T. Limmer\looseness=-1}
		\email{dlimmer@berkeley.edu}
		\affiliation{Department of Chemistry, University of California, Berkeley, CA 94720, USA\looseness=-1}
		\affiliation{Kavli Energy Nanoscience Institute at Berkeley, Berkeley, CA 94720, USA\looseness=-1}
		\affiliation{Chemical Sciences Division, Lawrence Berkeley National Laboratory, Berkeley, CA 94720, USA\looseness=-1}
		\affiliation{Materials Science Division, Lawrence Berkeley National Laboratory, Berkeley, CA 94720, USA\looseness=-1}


		\begin{abstract}
			In this paper we outline a physically motivated framework for describing spin-selective recombination processes in chiral systems, from which we derive spin-selective reaction operators for recombination reactions of donor-bridge-acceptor molecules, where the electron transfer is mediated by chirality and spin-orbit coupling. In general the recombination process is  selective only for spin-coherence between singlet and triplet states, and it is not in general selective for spin polarisation. We find that spin polarisation selectivity only arises in hopping mediated electron transfer. We describe how this effective spin-polarisation selectivity is a consequence of spin-polarisation generated transiently in the intermediate state. The recombination process also augments the coherent spin dynamics of the charge separated state, which is found to have a significant effect on recombination dynamics and to destroy any long-lived spin polarisation. Although we only consider a simple donor-bridge-acceptor system, the framework we present here can be straightforwardly extended to describe spin-selective recombination processes in more complex systems.
		\end{abstract}
		\maketitle
		\section{Introduction}
		
		There has recently been growing interest in spin-selectivity of processes in molecular chiral donor-acceptor systems.\cite{Aiello2022,Privitera2022,Evers2022,Volker2023} These systems may provide a versatile platform for exploring the molecular origins of the chirality induced spin selectivity (CISS) effect,\cite{Chiesa2021,Privitera2022a,Fay2021c,Luo2021} without the complications of interactions with electrodes,\cite{Evers2022,Naaman2020a,Waldeck2021,Aragones2017,Naskar2023} and it has also been suggested that chiral donor-acceptor systems could be exploited in various quantum information science applications.\cite{Aiello2022,Naaman2015,Wasielewski2023} Several theories have been proposed for the molecular CISS effect and how spin polarisation is generated in the formation of charge separated states,\cite{Fay2021,Fay2021c,Luo2021,Chandran2022,Chandran2022a} and a handful of experimental protocols have been proposed to test these theories.\cite{Fay2021,Luo2021,Chiesa2021,Privitera2022,Carmeli2014} Whilst most studies to date have focussed on the CISS effect in the formation of charge seperated states in donor-acceptor systems, it has also been suggested that CISS could play a role in the charge recombination of donor-acceptor systems.\cite{Fay2021,Luo2021} 
		However the proposed theories of CISS in charge recombination are either limited to simple one-step electron transfer\cite{Fay2021} or purely phenomenological,\cite{Luo2021} and thus far the role of CISS in experimentally realised donor-bridge-acceptor systems\cite{Junge2020,Wasielewski2006,Privitera2022,Mani2022} has not been explored. Motivated by this, in this paper we aim to answer the question: does chirality lead to spin-selective charge recombination in donor-bridge-acceptor molecules?  
		
		In systems where chirality and spin-orbit coupling have no effect on charge recombination of charge separated (CS) states, the treatment of spin-selective recombination is well understood.\cite{Haberkorn1976,Ivanov2010,Maeda2013,Fay2018} We describe the system with a time-dependent spin density operator for the CS state, $\op{\sigma}_{\mathrm{CS}}$, which spans the set of near-degenerate singlet and triplet spin states of the CS state. This spin density operator obeys the well-established Haberkorn quantum master equation
		\begin{align}
			\dv{t} \op{\sigma}_{\mathrm{CS}}(t) = -i[\op{H},\op{\sigma}_\mathrm{CS}(t)] + \pK \op{\sigma}_\mathrm{CS}(t)
		\end{align}
		where $\op{H}$ is the spin Hamiltonian for the CS state and $\pK$ is the reaction superoperator, which is given by
		\begin{align}
			\pK \op{\sigma}_\mathrm{CS} = -\left\{\op{K},\op{\sigma}_\mathrm{CS}\right\} = -\left\{\frac{k_{\mathrm{CR},\sing}}{2}\op{P}_\sing + \frac{k_{\mathrm{CR},\trip}}{2}\op{P}_\trip, \op{\sigma}_\mathrm{CS}\right\},
		\end{align}
		in which and $k_{\mathrm{CR},\sing}$ and $k_{\mathrm{CR},\trip}$ are the singlet and triplet spin selective reaction rate constants and $\op{P}_{\sing} = \dyad{\sing}$ and $\op{P}_{\trip} = \sum_{\alpha = x,y,z} \dyad{\trip_\alpha}$ are projection operators onto singlet and triplet spin states of the CS state.\cite{Haberkorn1976,Ivanov2010,Fay2018} $\pK$ describes the full effect of the reaction process on the dynamics of the CS state, and the reaction operator $\op{K}$ encodes how population is lost from the CS state. In what follows we will refer to $\pK$ as the reaction \emph{superoperator} and $\op{K}$ as the reaction \emph{operator}. It should be noted that this equation does not conserve the trace of $\op{\sigma}_{\mathrm{CS}}$ because population is lost from the CS state by recombination, and it further assumes that the recombination of the CS state is irreversible.
		
		It has been postulated by Luo \& Hore that in chiral donor-acceptor systems, the reaction operator $\op{K}$ for a CISS mediated recombination process should be given by\cite{Luo2021}
		\begin{align}\label{phen-K-eq}
			\op{K} = \frac{k_\mathrm{CR}}{2} \dyad{\phi_\chi}
		\end{align}
		in which $\ket{\phi_\chi} = \cos(\chi/2)\ket{\sing} + \sin(\chi/2)\ket{\trip_z}$, $\ket{\sing} = \frac{1}{\sqrt{2}}(\ket{\uparrow_\mathrm{D}\downarrow_\mathrm{A}}-\ket{\downarrow_\mathrm{D}\uparrow_\mathrm{A}})$ and $\ket{\trip_z} = \frac{1}{\sqrt{2}}(\ket{\uparrow_\mathrm{D}\downarrow_\mathrm{A}}+\ket{\downarrow_\mathrm{D}\uparrow_\mathrm{A}})$, where the quantisation axis, $z$, is the spin-polarisation axis, and the mixing angle $\chi$ is a phenomenological parameter that parametrises the extent of spin polarisation selectivity. This form of the reaction operator is based on the assumption that in a chiral molecule the recombination process is partially spin selective, and it is straightforward to verify that the total decay rate of the CS state with this model is dependent on the spin-polarisation $\ev{\Delta S_z} = \ev{S_{\mathrm{D}z}-S_{\mathrm{A}z}}$ of the CS state. The limit of full spin-selectivity is recovered for $\chi = \pi/2$ and $\ket{\phi_\chi} = \ket{\uparrow_\mathrm{D}\downarrow_\mathrm{A}}$, where the recombination only occurs if the donor and acceptor electrons have specific opposite spin orientations in the molecular frame. 
		It should be noted that this reaction operator has not been derived from any microscopic models of chirality induced spin selectivity, and there is no direct experimental evidence that it provides a reasonable model of spin selective recombination in chiral donor-acceptor systems. 
		
		In previous work we have derived a different reaction superoperator to that given by Eq.~\eqref{phen-K-eq} for a simple one-step electron transfer between a donor and an acceptor \ce{D^$\bullet+$-A^$\bullet-$ $\to$ D-A}, invoking the modest approximations.\cite{Fay2018} Specifically it is assumed that the coupling between the charge transfer states is weak, that the nuclear degrees of freedom are initially at local thermal equilibrium on one of the charge transfer potential energy surfaces, and that the Condon approximation holds for the direct charge transfer coupled, $V_{\mathrm{DA}}$, and the spin-orbit mediated charge transfer coupling $\Lambda_\mathrm{DA}$ (essentially the same approximations as Marcus theory\cite{May2000,Marcus1956,Hush1958}). The model electronic Hamiltonian for the singlet and triplet \ce{D^$\bullet+$-A^$\bullet-$} (CS) and \ce{D-A} ($\sing_0$) states in this theory is given by,
		\begin{align}\label{hda-eq}
			\begin{split}
				\op{H}_{\mathrm{DA}} &= E_{\sing_0}\dyad{\sing_0}+ \sum_{\Theta=\sing,\!\trip_x,\!\trip_y,\!\trip_z} E_{\mathrm{CS},\!\Theta} \dyad{\mathrm{CS},\!\Theta} \\
				&+ {V_{\mathrm{DA}}} \left(\dyad{\mathrm{CS},\sing}{\sing_0} +\dyad{\sing_0}{\mathrm{CS},\sing}\right) \\
&+ i \frac{\Lambda_{\mathrm{DA}}}{2}\left(\dyad{\mathrm{CS},\trip_z}{\sing_0} -\dyad{\sing_0}{\mathrm{CS},\trip_z}\right).
			\end{split}
		\end{align}
		and from this the reaction superoperator can be {derived} to be
		\begin{align}\label{one-step-K-eq}
			\pK \op{\sigma}_{\mathrm{CS}} = -\left\{\frac{k_{\mathrm{CR}}}{2}\dyad{\psi_\theta}{\psi_\theta},\op{\sigma}_{\mathrm{CS}} \right\} -\!i\!\left[\delta\epsilon\dyad{\psi_\theta}{\psi_\theta},\op{\sigma}_{\mathrm{CS}}\right]\!,
		\end{align}
		where $\ket{\psi_\theta} = \cos\theta \ket{\sing} + i \sin\theta \ket{\trip_z}$, with $\tan\theta = \Lambda_\mathrm{DA}/(2V_{\mathrm{DA}})$, where the quantization axis $z$ is defined by the spin-orbit coupling vector. This state is not spin polarised, i.e. $\ev{S_{\mathrm{D}z}} =\ev{S_{\mathrm{A}z}} = 0$, so in this case the recombination process is not selective for spin polarisation. However, the recombination process is selective for the imaginary part of the coherence between $\ket{\sing}$ and $\ket{\trip_z}$ states, $\op{\Pi}_{\sing\trip_z} =-i(\dyad{\sing}{\trip_z}-\dyad{\trip_z}{\sing})/2 $. 
		The shift term $\delta\epsilon \approx (V_{\mathrm{DA}}^2 + (\Lambda_{\mathrm{DA}}/2)^2)/(E_{\sing_0}-E_{\mathrm{CS}})$ appearing in this reaction superoperator is a superexchange mediated spin-orbit coupling interaction in the CS state that emerges as a result of the spin-orbit interaction between the CS state and the $\sing_0$ ground-state, which cannot in general be neglected.\cite{Fay2021,Fay2021c} 
		
		In subsequent work it was found that an interplay of spin-orbit coupling and exchange interactions in a two-step charge separation  following photo-excitation can produce spin polarisation, where a single-step electron transfer cannot.\cite{Fay2021c} This naturally raises the question: what role does chirality induced spin selectivity play in the charge \emph{recombination} of a donor-bridge-acceptor system? To address this question we will consider charge recombination in a chiral \ce{D-B-A} molecule and derive a reaction operator which describes the superexchange and incoherent hopping limits of the charge recombination process.
		
		\section{CISS in donor-bridge-acceptor charge recombination}
		\begin{figure}
			\includegraphics[width=0.48\textwidth]{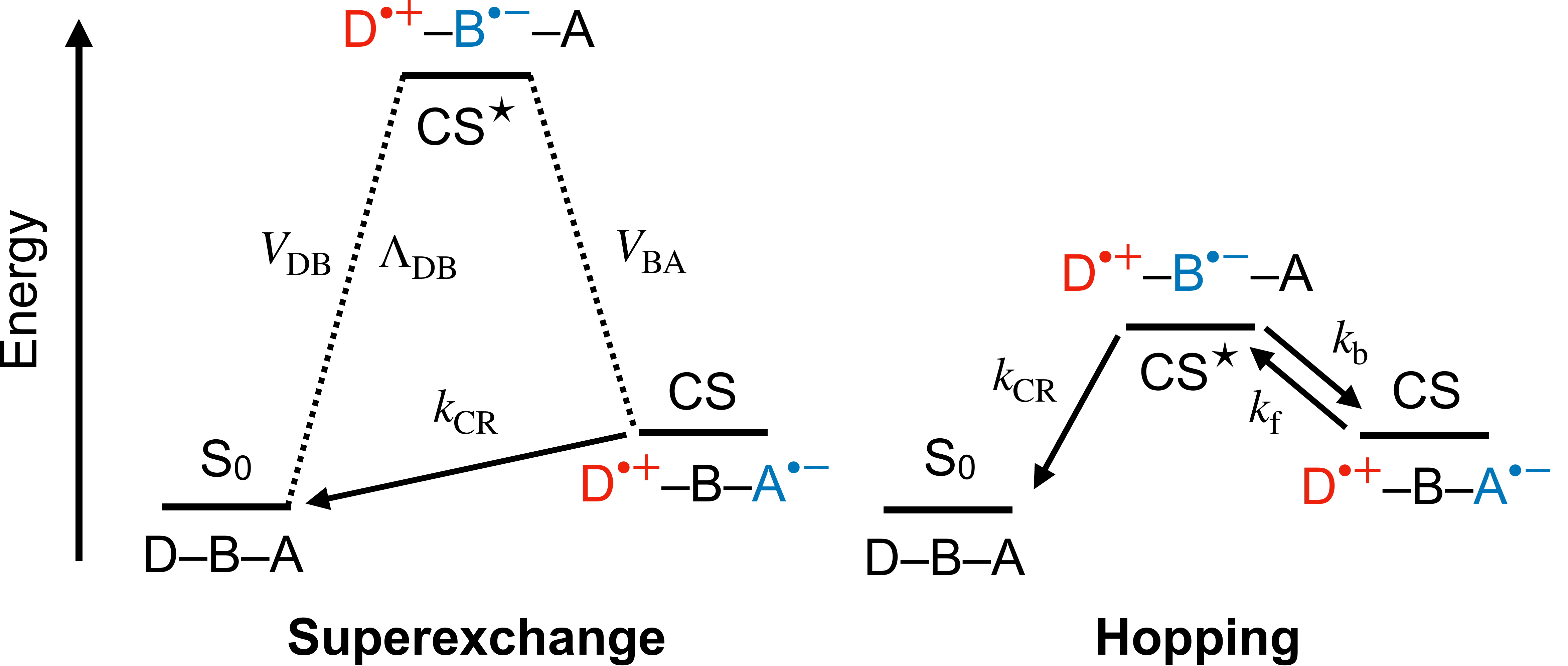}
			\caption{Diagram summarising the charge transfer states involved in the donor-bridge-acceptor system, indicating the incoherent rate processes with arrows and superexchange coupling with dashed lines, for the superexchange limit (left) and hopping limit (right).}\label{mechanism-fig}
		\end{figure}
		As a minimal model for a spin-orbit coupling mediated charge recombination process in a chiral donor-acceptor system, we consider recombination of a donor-bridge-acceptor system from a charge separated (CS) state, \ce{D^{$\bullet +$}-B-A^{$\bullet-$}}, back to a closed shell ground state ($\sing_0$), \ce{D-B-A}. We will also assume that direct \ce{A^{$\bullet -$}\to D^{$\bullet +$}} tunnelling of the electron does not occur, so the electron either hops via an intermediate \ce{CS^$\star$} state, \ce{D^{$\bullet +$}-B^{$\bullet-$}-A}, in which the bridge B is charged, or the electron tunnels via indirect superexchange coupling through virtual transitions to intermediate charge transfer states. Both the CS and $\ce{CS}^\star$ states can exist in either singlet or triplet electron spin states, whereas the $\sing_0$ ground state is assumed to only exist in a pure singlet state. For simplicity we will assume that only the \ce{B^{$\bullet -$}}$\to$\ce{D^{$\bullet$ +}} electron transfer is mediated by spin orbit coupling, and the initial hopping step is not spin selective. The transition from superexchange mediated tunnelling to hopping is controlled by the energy of the $\ce{CS}^\star$ state (at fixed electron transfer reorganisation energy). When the intermediate $\ce{CS}^\star$ state is very high in energy (compared to thermal energy $k_\mathrm{B} T$) hopping is unfeasible, thus superexchange dominates, but when the intermediate $\ce{CS}^\star$ is thermally accessible, hopping becomes the dominant mechanism.\cite{Hu1989,Petrov2001,May2000} This is illustrated schematically in Fig.~\ref{mechanism-fig}.\cite{Hu1989,Wasielewski2006}
		
		\subsection{The superexchange limit}
		
		Firstly, in order to understand the superexchange mediated electron transfer limit, we consider the following model for the electronic Hamiltonian for the $\sing_0$, \ce{CS} and $\ce{CS}^\star$ states,\cite{Fay2021c,Hu1989,May2000}
		\begin{align}\label{hfull-eq}
			\begin{split}
				\op{H}_{\mathrm{DBA}} &= E_{\sing_0} \dyad{\sing_0} +\!\!\!\!\! \sum_{\Theta=\sing,\!\trip_x,\!\trip_y,\!\trip_z} \!\!\!\!\!E_{\mathrm{CS}^\star,\Theta} \dyad{\mathrm{CS}^\star,\Theta} \\
				&+\!\!\!\!\! \sum_{\Theta=\sing,\!\trip_x,\!\trip_y,\!\trip_z} \!\!\!\!\!E_{\mathrm{CS},\Theta} \dyad{\mathrm{CS},\Theta} \\
				&+\!\!\!\!\! \sum_{\Theta=\sing,\!\trip_x,\!\trip_y,\!\trip_z} \!\!\!\!\!V_{\mathrm{BA}} \left(\dyad{\mathrm{CS}^\star,\Theta}{\mathrm{CS},\Theta} +\dyad{\mathrm{CS},\Theta}{\mathrm{CS}^\star,\Theta}\right) \\
				&+  V_{\mathrm{DB}} \left(\dyad{\mathrm{CS}^\star,\sing}{\sing_0} +\dyad{\sing_0}{\mathrm{CS}^\star,\sing}\right) \\
				&+ \frac{i\Lambda_{\mathrm{DB}}}{2} \left(\dyad{\mathrm{CS}^\star,\trip_z}{\sing_0} -\dyad{\sing_0}{\mathrm{CS}^\star,\trip_z}\right).
			\end{split} 
		\end{align}
		Here $\Theta = \sing, \trip_x,\trip_y$ or $\trip_z$, denotes the total electron spin state for the charge separated states, $V_{\mathrm{DB}}$ and $V_{\mathrm{BA}}$ denote spin-conserving diabatic (i.e. direct tunnelling) couplings between electronic states, and $\Lambda_{\mathrm{DB}}$ denotes the spin-orbit coupling between the $\ce{CS}^\star$ and $\sing_0$ ground state. For simplicty we assume that SOC only mediates the $\ce{D}\leftrightarrow\ce{B}$ coupling, but the general result we obtain holds when SOC mediates the $\ce{B}\leftrightarrow\ce{A}$ electron transfer process as well. 
		
		In the superexchange mediated electron transfer limit, where $E_{\mathrm{CS}^\star,\Theta} \gg E_{\mathrm{CS},\Theta}, E_{\sing_0}$, we do not need to explicitly include the intermediate charged bridge state, $\ce{CS}^\star$, and we can treat the charge recombination of the CS state to the \ce{S_0} state as a two electron transfer state problem, with an effective Hamiltonian given by
		\begin{align}\label{heff-se-eq}
			\begin{split}
			\op{H}_\mathrm{DBA,eff} &= \left(E_{\sing_0}\!+\!\delta E_{\sing_0}\right)\! \dyad{\sing_0}\!\\
			&+\!\!\!\!\!\!\!\!\! \sum_{\Theta=\sing,\!\trip_x,\!\trip_y,\!\trip_z}\!\!\!\!\!\!\!\!\!\! \left(E_{\mathrm{CS},\!\Theta} \!+\!\delta E_{\mathrm{CS},\Theta}\right)\!\dyad{\mathrm{CS},\!\Theta} \\
				&+ {V_{\mathrm{DA,eff}}} \left(\dyad{\mathrm{CS},\sing}{\sing_0} +\dyad{\sing_0}{\mathrm{CS},\sing}\right) \\
				&+ i \frac{\Lambda_{\mathrm{DA,eff}}}{2}\left(\dyad{\mathrm{CS},\trip_z}{\sing_0} -\dyad{\sing_0}{\mathrm{CS},\trip_z}\right).
			\end{split}
		\end{align}
		A derivation of this effective Hamiltonian is given in appendix \ref{heff-app}. We see from this that in the superexchange limit there is an effective spin-conserving diabatic coupling between the $\ce{S}_0$ and $\ce{CS}$ states given by $V_\mathrm{DA,eff} = {V_{\mathrm{DB}} V_{\mathrm{BA}}}/({\bar{E} - E_{\mathrm{CS}^\star,\sing}})$ as well as an effective spin-orbit interaction with coupling strength $\Lambda_\mathrm{DA,eff} = {\Lambda_{\mathrm{DB}} V_{\mathrm{BA}}}/({\bar{E} - E_{\mathrm{CS}^\star,\trip_z}})$, where $\bar{E}$ is the average energy of the $\sing_0$ and $\ce{CS}$ states.
		
		Because the electronic Hamiltonian reduces to an effective two-state model in the superexchange limit, the electron transfer can be regarded as occurring in a single step,
		$$\ce{D^{$\bullet +$}-B-A^{$\bullet-$}  ->[$k_\mathrm{CR}$] D-B-A}.$$
		The electron transfer rate $k_{\mathrm{CR}}$ is mediated by the spin-conserving superexchange coupling term $V_{\mathrm{DA,eff}}$, and the spin-orbit coupling mediated superexchange coupling ${\Lambda}_{\mathrm{DA,eff}}$, and (when the Condon approximation is applied to the interstate superexchange couplings\cite{May2000}) the theory reduces to that presented in Ref.~\onlinecite{Fay2021} and the recombination superoperator is given exactly by Eq.~\eqref{one-step-K-eq}, where $\Lambda_{\mathrm{DA}} = \Lambda_{\mathrm{DA,eff}}$ and $V_{\mathrm{DA}} = V_{\mathrm{DA,eff}}$. We see that in the superexchange limit there is no selectivity for spin-polarisation and only selectivity for spin-coherence.
		
		\subsection{The hopping limit}
		
		In the limit where charge recombination is controlled by hopping via an intermediate charge seperated state in which the bridge is charged,\cite{Hu1989} \ce{D^{$\bullet +$}-B^{$\bullet-$}-A} , which we denote \ce{CS^$\star$}. The kinetic scheme for this mechanism is simply
		$$\ce{D^{$\bullet +$}-B-A^{$\bullet-$} <=>[$k_\mathrm{f}$][$k_\mathrm{b}$] D^{$\bullet +$}-B^{$\bullet-$}-A ->[$k_\mathrm{CR}$] D-B-A}.$$
		The forward and backward hopping rates are given by $k_\mathrm{f}$ and $k_\mathrm{b}$ respectively, and the recombination from the $\ce{CS}^\star$ state is assumed to occur irreversibly at a rate $k_\mathrm{CR}$. We start by writing down coupled equations of motion for the spin density operators of the \ce{CS^$\star$} [\ce{D^{$\bullet +$}-B^{$\bullet-$}-A}] and \ce{CS} [\ce{D^{$\bullet +$}-B-A^{$\bullet-$}}] states, as derived in Refs.~\onlinecite{Fay2018,Fay2021,Fay2021c}.
		\begin{align}
			\begin{split}\label{sigma-csstar-eq}
			\dv{t}\op{\sigma}_{\mathrm{CS}^\star}(t) &= -\left\{\frac{k_{\mathrm{CR}}}{2}\dyad{\psi_\theta}{\psi_\theta},\op{\sigma}_{\mathrm{CS}^{\star}}(t) \right\} -k_\mathrm{b}\op{\sigma}_{\mathrm{CS}^\star}(t)\\
			   -i&[\delta\epsilon\dyad{\psi_\theta}{\psi_\theta}+2J \op{P}_\sing,\op{\sigma}_{\mathrm{CS}^{\star}}(t)] + k_\mathrm{f}\op{\sigma}_{\mathrm{CS}}(t)
			\\
		&= \pL_{\mathrm{CS}^\star} \op{\sigma}_{\mathrm{CS}^\star}(t)+ k_\mathrm{f}\op{\sigma}_{\mathrm{CS}^\star}(t) 
		\end{split}\\
		\dv{t}\op{\sigma}_{\mathrm{CS}}(t) &= -i[\op{H},\op{\sigma}_\mathrm{CS}(t)] + k_\mathrm{b}\op{\sigma}_{\mathrm{CS}^\star}(t) - k_\mathrm{f}\op{\sigma}_{\mathrm{CS}}(t),\label{sigma-cs-eq}
		\end{align}
		where $\pL_{\mathrm{CS}^\star} = -\left\{{k_{\mathrm{CR}}}/{2}\dyad{\psi_\theta}{\psi_\theta},\cdot \right\} -k_\mathrm{b}
		-i[\delta\epsilon\dyad{\psi_\theta}{\psi_\theta}+2J \op{P}_\sing,\cdot] $. Here we have assumed that the intermediate state is sufficiently short lived that we need only include the exchange interaction $J$ and the spin-orbit coupling shift term $\delta\epsilon$ in the spin Hamiltonian for the \ce{CS^$\star$} state, and we can safely neglect hyperfine, dipolar and Zeeman interaction terms. This equation assumes there is no long-lived coherence between different charge transfer states, and thus individual hopping steps can be treated as incoherent processes. We also assume the $\ce{CS}^\star$ spin states are near-degenerate (relative to thermal energy) so we do not need to account for spin-selectivity in the forward/backward hopping processes.\cite{Fay2021c}
		
		Before proceeding further we will outline a qualitatively how spin polarisation selectivity can emerge in hopping mediated charge recombination. When spin-orbit coupling mediates the charge recombination from the intermediate \ce{CS^$\star$} state, it will selectively remove the $\ket{\psi_\theta}$ state. Molecules in electron spin states orthogonal to $\ket{\psi_\theta}$, which have a non-zero spin-coherence, e.g. $\ket{\psi_{\theta\perp}} = \sin\theta \ket{\sing} - i \cos\theta \ket{\trip_z}$, 
		will not react and subsequently they will evolve coherently under the exchange interaction $J$ in the intermediate state to generate a spin polarised state.\cite{Fay2021c} 
		This spin polarisation is then transferred back to the CS state. In this sense we see that the effective loss of spin polarisation from the CS state is a result of the opposite spin polarisation being generated in the intermediate state in molecules which do not recombine, which is transferred back to the CS state.
		
		In order to derive the effective reaction superoperator, we apply the steady-state approximation to the \ce{CS^$\star$} spin density operator, $\dv{t}\op{\sigma}_{\mathrm{CS}^\star}(t) \approx 0$, from which we obtain the following equation for \ce{CS^$\star$} density operator in terms of the \ce{CS} density operator,
		\begin{align}\label{ssa-csstar-eq}
			\op{\sigma}_{\mathrm{CS}^\star}(t) \approx -k_\mathrm{f} \pL_{\mathrm{CS}^\star}^{-1}\op{\sigma}_{\mathrm{CS}^\star}(t)
		\end{align}
		and from this we can obtain the effective reaction superoperator as
		\begin{align}\label{ss-K-eq}
			\pK \op{\sigma}_{\mathrm{CS}}(t)  = - k_\mathrm{f}(1 + k_\mathrm{b} \pL_{\mathrm{CS}^\star}^{-1})\op{\sigma}_{\mathrm{CS}}(t).
		\end{align}
		Alternatively we can derive the hopping reaction superoperator without invoking the steady state approximation by first solving Eq.~\eqref{sigma-csstar-eq} for $\op{\sigma}_{\mathrm{CS}^{\star}}(t)$ to give
		\begin{align}
			\op{\sigma}_{\mathrm{CS}^{\star}}(t) = k_\mathrm{f}\int_0^t\dd{\tau} e^{\pL_{\mathrm{CS}^\star}\tau} \op{\sigma}_{\mathrm{CS}}(t-\tau) .
		\end{align}
		Assuming $e^{\pL_{\mathrm{CS}^\star}\tau}$ decays to zero on a time-scale faster than the dynamics of $\op{\sigma}_{\mathrm{CS}}(t-\tau)$, we can invoke a Markovian approximation where we replace $\op{\sigma}_{\mathrm{CS}}(t-\tau)\to \op{\sigma}_{\mathrm{CS}}(t)$ in the integral and we set the upper limit of the integral to $\tau = \infty$. With this we arrive at the same result as the steady-state approximation for $\op{\sigma}_{\mathrm{CS}^{\star}}(t) $ given by Eq.~\eqref{ssa-csstar-eq}, but this approach shows that the SSA approximation can be expected to be accurate provided the spin dynamics of the $\ce{CS}^\star$ occur on a much faster time scale than those of the $\ce{CS}$ state. 
		Unfortunately this reaction superoperator does not reduce to a simple form in the general case, although an analytical expression can be obtained. However we can examine particular limits and extract the effective reaction operator $\op{K}$.
		
		Firstly it is instructive to consider the case where $\theta =0$ and $\delta\epsilon = 0$, i.e. when there is no spin-orbit coupling involved in the charge recombination process and there is no CISS effect. In this case the reaction operator reduces to
		\begin{subequations}\label{hopping-eqs}
			\begin{align}
				\begin{split}
					\pK \op{\sigma}_{\mathrm{CS}}  &= -\left\{\frac{\tilde{k}_{\mathrm{CR}}}{2}\op{P}_\sing,\op{\sigma}_{\mathrm{CS}}\right\} -i \left[2\tilde{J}\op{P}_\sing , \op{\sigma}_{\mathrm{CS}}\right]\\
					&+\tilde{k}_\mathrm{D} \left(\op{P}_\sing\op{\sigma}_{\mathrm{CS}}\op{P}_\sing - \frac{1}{2}\left\{\op{P}_\sing,\op{\sigma}_{\mathrm{CS}}\right\}  \right)
				\end{split} 
			\end{align}
		\end{subequations}
		where $\tilde{k}_\mathrm{CR} = k_\mathrm{f}k_\mathrm{b}/(k_\mathrm{b} + k_\mathrm{CR})$ is the effective reaction rate, $2\tilde{J} = 8 J k_\mathrm{b} k_\mathrm{f}/[(4J)^2 + (k_\mathrm{CR} + 2 k_\mathrm{b})^2]$ is an effective exchange coupling, and $\tilde{k}_\mathrm{D} = k_\mathrm{f}((4J)^2 + k_\mathrm{CR}(k_\mathrm{CR} + 2 k_\mathrm{b}))/[(4J)^2 + (k_\mathrm{CR} + 2 k_\mathrm{b})^2]- \tilde{k}_\mathrm{CR}/2$, is an effective singlet-triplet dephasing rate. 
		In the limit where $J=0$, we can similarly obtain a simple expression for the reaction operator
		\begin{align}\label{J0-K-eq}
			\begin{split}
				\pK \op{\sigma}_{\mathrm{CS}}  &= -\left\{\frac{\tilde{k}_{\mathrm{CR}}}{2}\dyad{\psi_\theta},\op{\sigma}_{\mathrm{CS}}\right\} -i \left[\delta\tilde{\epsilon}\dyad{\psi_\theta} , \op{\sigma}_{\mathrm{CS}}\right]\\
				&+\tilde{k}_\mathrm{D} \left(\dyad{\psi_\theta}\op{\sigma}_{\mathrm{CS}}\dyad{\psi_\theta} - \frac{1}{2}\left\{\dyad{\psi_\theta},\op{\sigma}_{\mathrm{CS}}\right\}  \right)
			\end{split} 
		\end{align}
		where the effective charge recombination rate $\tilde{k}_\mathrm{CR} $, decoherence rate $\tilde{k}_\mathrm{D}$, and spin-orbit interaction $\delta\tilde{\epsilon}$, are given by the above expressions with the simple replacement $2J\to\delta\epsilon$. 
		In each case we see that the effective recombination operator can be decomposed into a shift in the spin Hamiltonian $\delta \op{H}$, a Lindbladian decoherence term, with Lindblad rates and operators $\gamma_j$ and $\op{L}_j$, and a reaction term with the reaction operator $\op{K}$,
		\begin{align}
			\begin{split}
			\pK \op{\sigma}_{\mathrm{CS}} &= -\left\{\op{K},\op{\sigma}_{\mathrm{CS}}\right\} - i \left[\delta \op{H},\op{\sigma}_{\mathrm{CS}}\right] \\
			& + \sum_{j}\gamma_j \left(\op{L}_j\op{\sigma}_{\mathrm{CS}} \op{L}_j^\dag - \frac{1}{2}\left\{\op{L}_j^\dag \op{L}_j , \op{\sigma}_{\mathrm{CS}} \right\} \right).
			\end{split}
		\end{align}
		Firstly we will consider the reaction operator $\op{K}$, although the decoherence and Hamiltonian shift terms cannot in general be neglected and we will later evaluate their importance. In general $\op{K}$ can be decomposed into the following set of operators
		\begin{align}\label{K-gen-eq}
			\op{K} = \frac{\tilde{k}_\sing}{2}\op{P}_\sing + \frac{\tilde{k}_{\trip_z}}{2}\op{P}_{\trip_z} + \frac{\tilde{k}_z}{2} \Delta\op{S}_z + \frac{\tilde{k}_{\sing\trip_z}}{2} \op{\Pi}_{\sing\trip_z}
		\end{align}
		where $\op{P}_{\trip_z} = \dyad{\trip_z}$ is a projection operator onto the $\trip_z$ state, $\Delta \op{S}_z = (\dyad{\sing}{\trip_z}+\dyad{\trip_z}{\sing})/2 = \dyad{\uparrow_\mathrm{D}\downarrow_\mathrm{A}}-\dyad{\downarrow_\mathrm{D}\uparrow_\mathrm{A}}$ is the spin polarisation operator and $\op{\Pi}_{\sing\trip_z} =-i(\dyad{\sing}{\trip_z}-\dyad{\trip_z}{\sing})/2 $ is the spin coherence operator. In general the rates must satisfy $\tilde{k}_\sing,\tilde{k}_{\trip_z} \geq 0$ and $
		\tilde{k}_{\sing}\tilde{k}_{\trip_z} \geq (\tilde{k}_{\sing\trip_z}^2 + \tilde{k}_z^2)/4$ in order to preserve positivity of the density operator. We note that in the case of Luo \& Hore's phenomenological theory, we have $\tilde{k}_\sing = k_{\mathrm{CR}}\cos^2(\chi/2)$, $\tilde{k}_{\trip_z} = k_{\mathrm{CR}}\sin^2(\chi/2)$, $\tilde{k}_z = k_{\mathrm{CR}} \sin(\chi)$ and $\tilde{k}_{\sing\trip_z} = 0$.\cite{Luo2021}
		
		We find in general that both $\tilde{k}_z$ and $\tilde{k}_{\sing\trip_z}$ are proportional to $\sin(2\theta)$, meaning they only emerge in chiral systems. This is because the spin coherence that is generated by the spin-orbit mediated recombination is proportional to $\sin(2\theta)$. Furthermore we find that the ratio of $\tilde{k}_z$ to $\tilde{k}_{\sing\trip_z}$ is given by
		\begin{align}
			\frac{\tilde{k}_z}{\tilde{k}_{\sing\trip_z}} = \frac{4J (2 k_\mathrm{b} + k_\mathrm{CR})}{ (2 k_\mathrm{b} + k_\mathrm{CR})^2 + 4 \delta\epsilon^2 + 8J\delta\epsilon\cos 2\theta}
		\end{align}
		which shows that spin polarisation selectivity can only arise if $J$ is non-zero, and if the back reaction and charge recombination from the intermediate state are sufficiently slow to allow some degree of coherent spin evolution in the intermediate state to generate spin polarisation.
		
		The full expressions for the spin-selective rate constants are somewhat cumbersome, although straightforward to evaluate numerically, 
		so here we will analyse some of their properties in specific limits. In appendix \ref{hopping-app} we show how to obtain $\pK$ in the weak spin-orbit coupling limit, which we expect to be applicable to many organic donor-bridge-acceptor systems, where the reaction superoperator parameters are relatively simple. In this limit we find there are two dominant decoherence processes, with Lindblad operators $\op{L}_j = \op{P}_\sing$ and $\op{P}_{\trip_z}$, and the recombination process is selective for both the $\sing$ and $\trip_z$ states, as well as the spin polarisation and $\sing$-$\trip_z$ coherence $\op{\Pi}_{\sing\trip_z}$. We also find that the spin-coherence and spin-polarisation selectivity only emerge at first-order in the spin orbit coupling, and spin-polarisation selectivity is only non-zero if $J$ is non-zero.
		
		Starting from the weak spin-orbit coupling limit results in appendix \ref{hopping-app}, it is instructive to consider the limit where coherent dynamics of the \ce{CS^$\star$} spins are much slower than the incoherent recombination process, i.e. $k_\mathrm{b},k_\mathrm{CR} \gg J,\delta\epsilon$. In this limit we find
\begin{subequations}
\begin{align}
	{\tilde{k}_{\sing}} &\approx {\cos^2\theta}\frac{k_\mathrm{f}k_\mathrm{CR}}{k_\mathrm{CR}+k_\mathrm{b}} , \
	{\tilde{k}_{\trip_z}}\approx {\sin^2\theta}  \frac{k_\mathrm{f}k_\mathrm{CR}}{k_\mathrm{b}} \\
	{\tilde{k}_{\sing\trip_z}} &\approx {\sin(2\theta)}\frac{k_\mathrm{f}k_\mathrm{CR}}{k_\mathrm{CR}+k_\mathrm{b}} , \ 
	{\tilde{k}_z} \approx \frac{k_\mathrm{f}k_\mathrm{CR}}{k_\mathrm{CR}+k_\mathrm{b}} \frac{4 J \sin2\theta}{2 k_\mathrm{b} + k_\mathrm{CR}}.
\end{align}
\end{subequations}		
		We see that the spin polarisation removed from the CS state is proportional to $J$ in the intermediate state. 
		Conversely, in the limit where $J \gg k_\mathrm{b},k_\mathrm{CR} ,\delta\epsilon$, it can be found that $\tilde{k}_z \approx 0 $ and $\tilde{k}_{\sing\trip_z}\approx 0$. This is because in the large $J$ limit spin polarisation generated by the exchange interaction oscillates many times in the intermediate state prior to transfer back to the CS state, which averages the spin polarisation that is transferred back to zero.
		
		It is also interesting to consider the limit where $k_\mathrm{CR}$ is small and $\delta\epsilon = 0$. In this case we find
		\begin{subequations}
		\begin{align}
			{\tilde{k}_{\sing}} &\approx {\cos^2\theta}\frac{k_\mathrm{f}k_\mathrm{CR}}{k_\mathrm{b}} , \
			{\tilde{k}_{\trip_z}}\approx {\sin^2\theta}  \frac{k_\mathrm{f}k_\mathrm{CR}}{k_\mathrm{b}} \\
			{\tilde{k}_{\sing\trip_z}} &\approx \frac{k_\mathrm{f}k_\mathrm{CR} k_\mathrm{b} \sin2\theta}{k_\mathrm{b}^2 + (2J)^2} , \
			{\tilde{k}_z} \approx  \frac{(2 J)k_\mathrm{f}k_\mathrm{CR}\sin2\theta}{k_\mathrm{b}^2 + (2J)^2}.
		\end{align}
		\end{subequations}
		In this case, if the initial charge separation goes via the same intermediate \ce{CS^$\star$} state and if the initial charge separation following photo-excitation forms the \ce{CS^$\star$} state in the $\ket{\psi_\theta}$ state, i.e. assuming $\Lambda_\mathrm{D^\star B}/V_\mathrm{D^\star B} = \Lambda_\mathrm{DB}/V_\mathrm{DB}$ (where $\ce{D}^\star$ denotes the excited precursor donor orbital), then the spin polarisation in the initial CS state is $-{(2 J)k_\mathrm{b} \sin2\theta}/({k_\mathrm{b}^2 + (2J)^2})$ and the initial spin coherence is ${k_\mathrm{b}^2 \sin2\theta}/({k_\mathrm{b}^2 + (2J)^2})$. So in this limit the charge recombination is selective for the same spin-coherence as is initially generated but the \emph{opposite} spin polarisation. It is straightforward to show that this result also holds in the case where $\delta\epsilon \neq 0$. It should be noted that the situation is more complicated if the phase $\Lambda_\mathrm{D^\star B}/V_\mathrm{D^\star B}$ is different to $\Lambda_\mathrm{DB}/V_\mathrm{DB}$, which could be the case in a real system. 
		
		We now turn to the shift term, $\delta \op{H}$, in the full effective reaction superoperator, which arises from coherent dynamics which occur transiently in the intermediate state. We can expand this shift term in terms of the operators $\op{P}_{\sing},\ \op{P}_{\trip_z}$, $\Delta\op{S}_z$ and $\op{\Pi}_{\sing\trip_z}$, as we did for $\op{K}$,
		\begin{align}
			\delta\op{H} = \delta\tilde{\epsilon}_\sing \op{P}_{\sing}+ \delta\tilde{\epsilon}_{\trip_z} \op{P}_{\trip_z} +\delta\tilde{\epsilon}_z \Delta\op{S}_z +\delta\tilde{\epsilon}_{\sing\trip_z}  \op{\Pi}_{\sing\trip_z}.
		\end{align}
		Evaluating the reaction superoperator, we find in general that $\delta\tilde{\epsilon}_{z} = 0$, whilst the other terms are non-zero. The non-zero terms account for the shift in energy of the singlet and triplet states due to the exchange coupling in the intermediate state, and an effective spin-orbit coupling that arises due to the superexchange spin-orbit interaction in the \ce{CS^$\star$} state (the term proportional to $\delta \epsilon$ in Eq.~\eqref{sigma-csstar-eq}). As with $\op{K}$ it is possible to obtain a exact expression for $\delta \op{H}$, but it is very complicated, however in the weak spin-orbit coupling limit (presented in appendix \ref{hopping-app}) it is relatively straightforward to evaluate. The presence of this shift term (and the decoherence terms) couples and subsequently mixes spin polarised states of the CS state, which means there exists no ``protected'' spin polarised CS state that can be generated from an non-polarised initial spin state. Importantly this implies there can be no long-lived spin polarisation in the CS state. In numerical tests presented in the next section, we will show that this shift term is essential in accurately calculating the spin polarisation and coherence dynamics of the CS state. 
		\begin{figure}
			\includegraphics[width=0.425\textwidth]{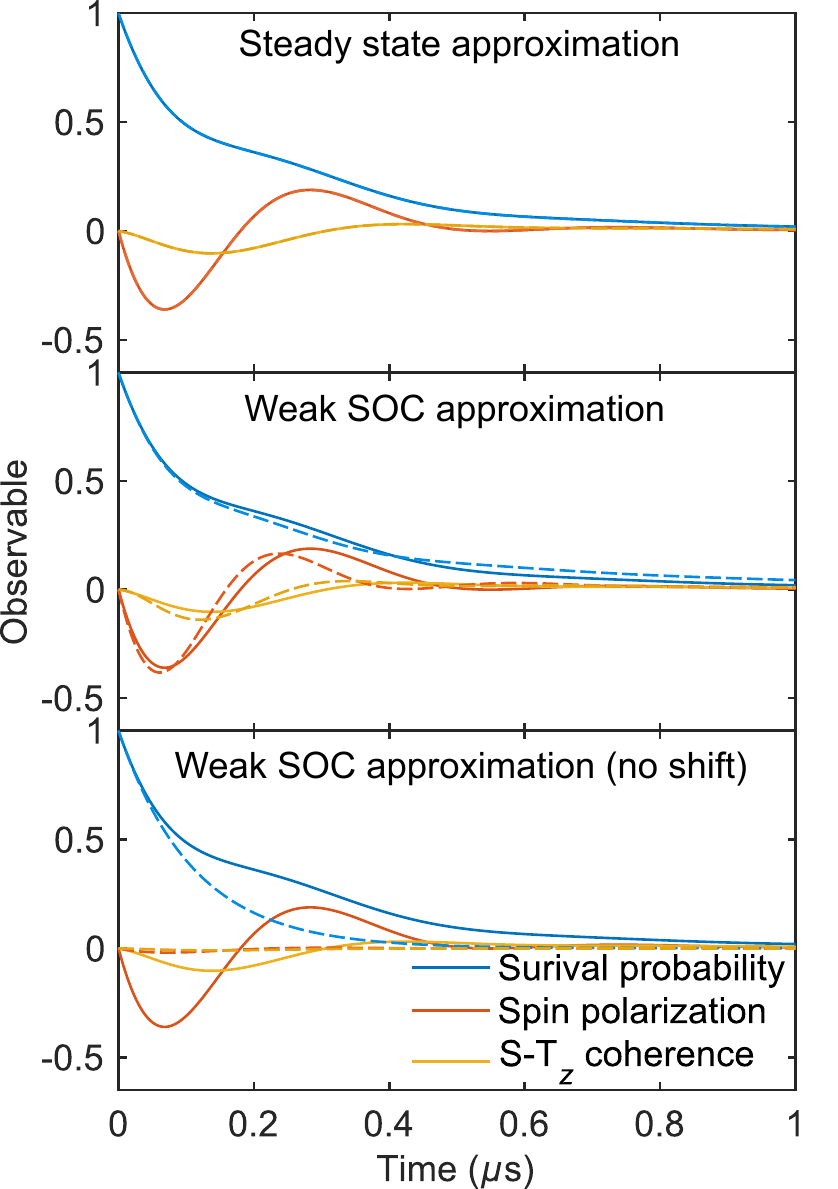}
			\caption{Dynamics of the hopping model calculated with the full set of density operator equations (solid lines) and various levels of approximation for the reaction superoperator (dashed lines). In this example $J = 1000$ mT, $\delta \epsilon = - 0.98 J$, $k_\mathrm{f} = 100\ \mu\mathrm{s}^{-1}$, $k_\mathrm{b} = 100 k_\mathrm{f}$, and $k_{\mathrm{CR} }= 10 k_\mathrm{f}$. }\label{numerics-fig}
		\end{figure}
	
		\section{Numerical tests for the hopping limit}
	
		In this section we aim to evaluate the accuracy of the steady state approximation used in the previous section to obtain the reaction superoperator, as well as the relative importance of spin selectivity in the reaction operator $\op{K}$ and augmented spin dynamics generated by $\delta \op{H}$. As a first test, in Fig.~\ref{numerics-fig} we show the dynamics of CS state population and the spin polarization in the CS state as a function of time for the CS state, for a model chiral \ce{D-B-A} system undergoing hopping mediated charge recombination, where the intermediate $\ce{CS}^\star$ state has a lifetime of approximately 10 ns (the complete set of model parameters are given in the figure caption). For this example we can calculate the spin dynamics with the CS state initialized in the singlet spin state ($\op{\sigma}_{\mathrm{CS}}(t = 0) = \dyad{\sing}$) for the full model, Eqs.~\eqref{sigma-csstar-eq} and \eqref{sigma-cs-eq}, for the full steady state approximation, Eq.~\eqref{ss-K-eq}, and for the weak spin-orbit coupling reaction superoperator obtained in appendix \ref{hopping-app}, and neglecting the weak spin-orbit coupling contribution to the $\delta \op{H}$ (i.e. setting $\delta \epsilon_{\sing\trip_z}^{(1)}$ in Eq.~\eqref{dH1-eq} to zero). 
		
		In the top panel of Fig.~\ref{numerics-fig} we show  the CS survival probability (blue), $p(t) = \Tr[\op{\sigma}_{\mathrm{CS}}(t)]$, and spin polarization (red), $\ev{\Delta S_z}$, and spin coherence (gold), $\ev{\Pi_{\sing\trip_z}}$, for the full model, Eqs.~\eqref{sigma-csstar-eq} and \eqref{sigma-cs-eq}], (solid lines) and for the full effective reaction superoperator, Eq.~\eqref{ss-K-eq}, (dashed lines). In this example both spin coherence and spin polarisation are generated transiently by the hopping process, and oscillations in these quantities induce oscillations in the CS state decay rate. We also see that the net spin-polarisation generated in the CS state eventually decays to zero. This is because all spin-polarised states are coupled by effective interactions in $\delta \op{H}$, which means the CS state fully decays, leaving zero net spin-polarisation. We see the full model result and effective reaction superoperator obtained with the steady-state approximation agree to graphical accuracy, even in this model with physically reasonable parameters where the frequency associated with $2J$ is significantly faster than the decay rate of the intermediate state. In the middle panel we compare the full model (solid lines) and the weak spin-orbit coupling reaction superoperator, Eq.~\eqref{wsoc-K-eq} (dashed lines), which recovers most of the population and spin-polarisation dynamics, but is not quantitatively accurate for this example. In the bottom panel of Fig.~\ref{numerics-fig} we show results for the same calculation neglecting $\pazocal{O}(\sin\theta)$ contributions to $\delta \op{H}$ in the weak spin-orbit coupling reaction superoperator, Eq.~\ref{wsoc-K-eq}. This approximation fails to capture the spin polarisation dynamics accurately in this example, which demonstrates that the emergent spin polarisation in the CS state is primarily generated by augmentation of the coherent spin dynamics, rather than spin-selective recombination. 
		
		\begin{figure}
			\includegraphics[width=0.425\textwidth]{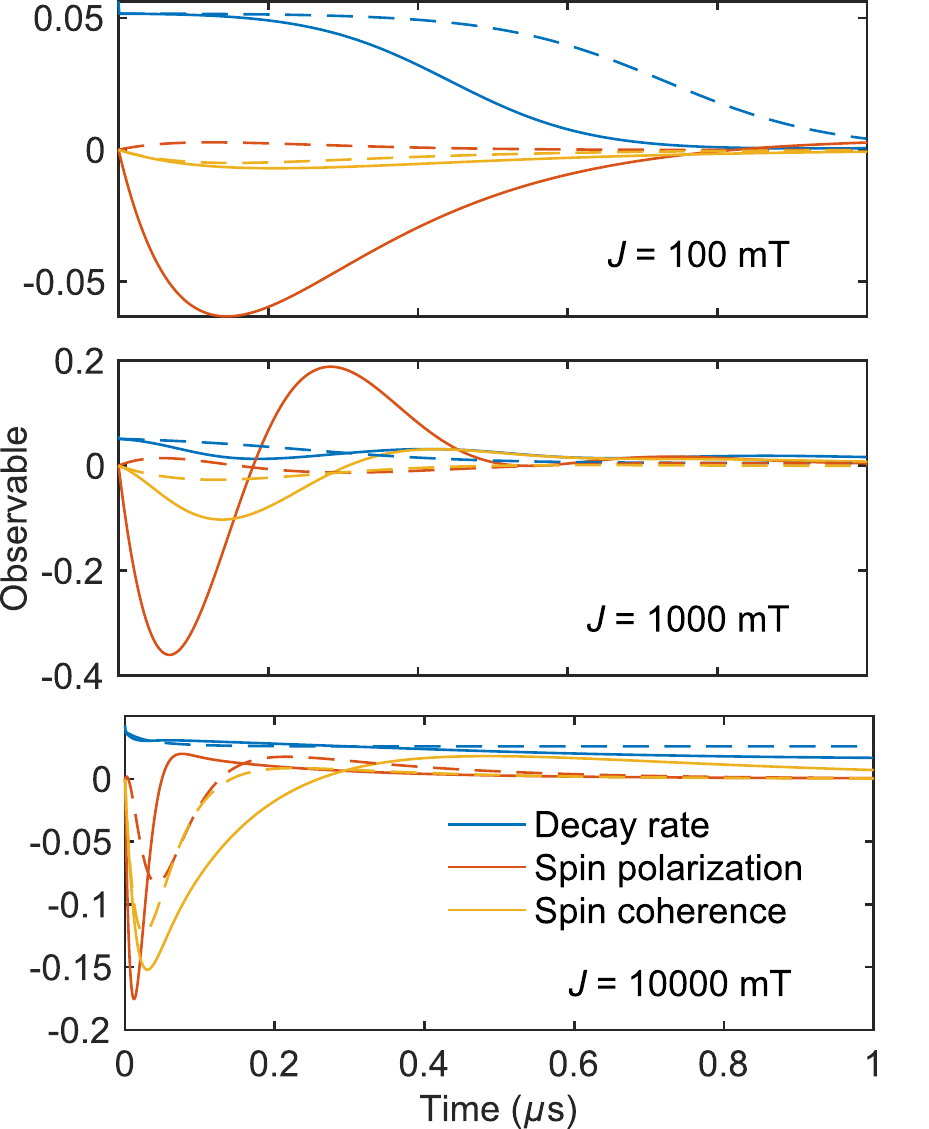}.
			\caption{Dynamics of the hopping model calculated with the full set of density operator equations (solid lines) and the full steady state reaction superoperator with $\delta \op{H} = 0$ (dashed lines). In this example $J $ is varied and in each case $\delta \epsilon = - 0.98 J$, $k_\mathrm{f} = 100\ \mu\mathrm{s}^{-1}$, $k_\mathrm{b} = 100 k_\mathrm{f}$, and $k_{\mathrm{CR} }= 10 k_\mathrm{f}$. }\label{heff-fig}
		\end{figure}
		To further test the importance of $\delta \op{H}$ we have calculated the decay rate, $k(t) = -\dot{p}(t) / p(t)$, spin polarization and spin coherence for the same model for a range exchange couplings $J$ and $\delta\epsilon$ values in the intermediate state, as shown in Fig.~\ref{heff-fig}. In each case we calculate the full spin dynamics of the CS state (solid lines), which agree to graphical accuracy with the full steady-state approximation results, and compare this to the dynamics with the steady-state approximation reaction superoperator with $\delta \op{H} = 0$ (dashed lines). We see that across a range of $J$ values (with $\delta \epsilon = -0.98 J$ in each case) that $\delta \op{H}$ cannot be neglected, even when the full stead-state approximation for the reaction operator and decoherence term are used. This approximation can even be qualitatively wrong, for example when $\delta \op{H}$ is neglected in the $J = 100$ mT and $J = 1000$ mT examples the sign of the predicted spin polarization is wrong. We also see that even when $\delta \op{H}$ is set to zero, decoherence contributions to $\pK$ destroy long-lived spin polarisation. Overall these tests show that both the reaction operator and the Hamiltonian shift play a significant role in determining the spin polarisation dynamics and survival probability of the CS state in this example, and the Hamiltonian shift $\delta \op{H}$ cannot simply be ignored, so the full reaction superoperator should always be used in calculations.

		\section{Conclusions}
		
		In this work we have derived a description of spin-selective electron transfer in chiral donor-bridge-accpetor systems through the reaction superoperator formalism. We have obtained expressions for this superoperator applicable in both the superexchange and hopping regimes for the recombination process. The form of the reaction superoperator is very simple in the superexchange limit, where it is not selective for spin polarisation. However in the hopping mediated limit the reaction superoperator becomes more complicated and we have found the recombination is selective for both spin-polarisation and spin-coherence in this case. The spin polarisation selectivity should be understood as arising from spin-polarisation being generated in molecules initially in non-reactive spin states, which is transferred back to the CS state in reverse-hopping. We also find that spin-polarisation selectivity emerges only when the intermediate charge separated state is sufficiently long lived, and when an exchange coupling in this state is large enough to generate spin polarisation in molecules in non-reactive spin states. Spin coherence selectivity and spin-polarisation selectivity can differ from the chirality induced spin coherence and polarisation generated by photo-excitation, and the selectivity depends on the phases couplings between bridge orbitals and the orbitals in the ground state and the excited precursor. This inverse spin polarisation selectivity of formation recombination could have some biological function, for example in magnetoreception\cite{Luo2021,Tiwari2022,Tiwari2022a} or in hindering reverse electron transfer in photosynthetic reaction centres.\cite{Carmeli2014,Naaman2022}
		
		Numerical tests have shown that in hopping mediated charge recombination chirality dependent shifts in the spin Hamiltonian, which are induced by transient dynamics in the intermediate state, also play an essential role in determining the survival probability and spin polarisation in the charge separated state. This shift Hamiltonian mixes all spin-polarised states, thereby destroying long-lived spin polarisation in the CS state. In order to accurately describe dynamics of charge recombination in chiral molecules it is clearly necessary to account for both spin-selectivity in the reaction and augmentation of the coherent dynamics by the recombination process. It is important to note that the theory proposed here is not fully consistent with the phenomenological treatments of CISS which have been proposed recently,\cite{Luo2021} and which have been used investigate the role of CISS in avian magnetoreception.\cite{Luo2021,Tiwari2022,Tiwari2022a} Although the theory we have presented does not have the simplicity of this phenomenological treatment, it is derived from a physically reasonable model, and all parameters appearing in the theory (such as forward and backward hopping rates) can in principle be measured experimentally or calculated using computational approaches. Experimental and computational studies suggest hopping mediated charge recombination may play a role in avian magnetoreception,\cite{Xu2021,Wong2021} and with the theory presented here it would be possible to rigorously study chirality-mediated spin effects in magnetoreception. The spin density operator framework used here can be extended straightforwardly to include other important physics of spin-correlated charge separated states necessary to understand real systems, such as hyperfine coupling effects and spin relaxation.\cite{Steiner1989,Nicholas2010,Chiesa2021,Fay2019b,Keens2020,Fay2021b} We anticipate that this framework for understanding spin selective charge recombination in chiral systems will be useful in numerous contexts involving molecular CISS, such as in devising systems exploiting CISS for quantum information science and in studies of CISS effects in biological electron transfer.

		\section*{Acknowledgements}
		T. P. F and D.T.L. were supported by the U.S. Department of Energy, Office of Science, Basic Energy Sciences, CPIMS Program Early Career Research Program under Award DE-FOA0002019.

	\appendix 
	
	\section{Effective Hamiltonian theory for superexchange electron transfer}\label{heff-app}
	
	
	Here we derive the effective Hamiltonian Eq.~\eqref{heff-se-eq} starting from the full model electronic state Hamiltonian given by Eq.~\eqref{hfull-eq}. We assume that the couplings $\Lambda_{\mathrm{DB}}$, $V_\mathrm{DB}$ and $V_{\mathrm{BA}}$ are small and that $E_\mathrm{CS^\star}\gg E_{\sing_0},E_{\mathrm{CS},\Theta}$. In this case the eigenvectors of $\op{H} = \op{H}_\mathrm{DBA} $ can be partitioned into two approximate subspaces, one spanned by the
	high energy $\ce{CS}^\star$ states and one spanned by the lower energy $\ce{CS}$ and $\sing_0$ states. With this observation, we can derive an approximate form for the electronic Hamiltonian within lower energy subspace. We first define a projection operator onto the low energy subspace $\op{P} = \dyad{\sing_0} + \sum_{\Theta}\dyad{\ce{CS},\Theta}$, and its complement $\op{Q} = 1 - \op{P}$. We project the electronic energy eigenstate equation $\op{H}\ket{\Psi} = E\ket{\Psi}$ to obtain equations for $\ket{\Psi}$ in the $\op{P}$ and $\op{Q}$ projected spaces,\cite{Mila2011}
	\begin{align}
		E\op{P}\ket{\Psi} &= \op{P}\op{H}\op{P}\ket{\Psi} + \op{P}\op{H}\op{Q}\ket{\Psi} \\
		E\op{Q}\ket{\Psi} &= \op{Q}\op{H}\op{P}\ket{\Psi} + \op{Q}\op{H}\op{Q}\ket{\Psi}.
	\end{align}
	Solving the equation for $\op{Q}\ket{\Psi}$, and substituting this into the equation for $\op{P}\ket{\Psi}$ yields an effective equation for $\op{P}\ket{\Psi}$
	\begin{align}
		E\op{P}\ket{\Psi} = [\op{P}\op{H}\op{P} + \op{P}\op{H}\op{Q}(E-\op{Q}\op{H}\op{Q})^{-1}\op{Q}\op{H}\op{P}]\op{P}\ket{\Psi}.
	\end{align}
	The effective Hamiltonian appearing on the right-hand side of this equation is dependent on E, but by exploiting the large separation between the low and high energy states, we can replace $(E-\op{Q}\op{H}\op{Q})^{-1} \to (\bar{E}-\op{Q}\op{H}\op{Q})^{-1}$ where $\bar{E}$ is the mean energy of the CS and $\sing_0$ states. With this approximation and by noting that $\op{P}\op{H}\op{P}$, $\op{Q}\op{H}\op{Q}$ and $\op{P}\op{H}\op{Q}$, with $\op{H} = \op{H}_\mathrm{DBA}$, can be written as
	\begin{align}
		\op{P}\op{H}\op{P} &= E_{\sing_0} \dyad{\sing_0} + \sum_{\Theta} E_{\mathrm{CS},\Theta} \dyad{\mathrm{CS},\Theta} \\
		\op{Q}\op{H}\op{Q} &= \sum_{\Theta} E_{\mathrm{CS}^\star,\Theta} \dyad{\mathrm{CS}^\star,\Theta} \\
		\begin{split}
			\op{P}\op{H}\op{Q} &= \sum_{\Theta} V_{\mathrm{BA}} \dyad{\mathrm{CS},\Theta}{\mathrm{CS}^\star,\Theta} \\
			&+  V_{\mathrm{DB}} \dyad{\sing_0}{\mathrm{CS}^\star,\sing} -\frac{i\Lambda_{\mathrm{DB}}}{2} \dyad{\sing_0}{\mathrm{CS}^\star,\trip_z}
		\end{split}\\
		\op{Q}\op{H}\op{P} &=(\op{P}\op{H}\op{Q})^\dag,
	\end{align}
	we straightforwardly obtain Eq.~\eqref{heff-se-eq} for $\op{H}_\mathrm{DBA,eff}$, where the energy level shifts in Eq.~\eqref{heff-se-eq} are given by,
	\begin{align}
		\delta E_{\sing_0} &= \frac{V_\mathrm{DB}^2}{\bar{E}-E_{\mathrm{CS}^\star,\sing}}+\frac{\Lambda_\mathrm{DB}^2}{4(\bar{E}-E_{\mathrm{CS}^\star,\trip_z})} \\
		\delta E_{\ce{CS},\Theta} &= \frac{V_\mathrm{BA}^2}{\bar{E}-E_{\mathrm{CS}^\star,\Theta}} .
	\end{align}

	\section{The weak SOC reaction operator}\label{hopping-app}
	
	In order to derive the weak spin-orbit coupling reaction superoperator, we partition $\pL_{\mathrm{CS}^\star} $ into a sum of a reference term, $\pL_\mathrm{d}$ which is diagonal in the singlet-triplet basis, and a term which couples singlet and triplet states, $\pL_\mathrm{c}$. The diagonal term is given by
	\begin{align}
		\pL_\mathrm{d} = -\left\{\frac{{k}_\sing}{2}\op{P}_\sing +\frac{{k}_{\trip_z}}{2}\op{P}_{\trip_z} + \frac{k_\mathrm{b}}{2},\ \cdot\  \right\} - i [\epsilon_\sing\op{P}_\sing +\epsilon_{\trip_z}\op{P}_{\trip_z}, \ \cdot\  ]
	\end{align} 
	where the effective rate constants and energies are given by
	\begin{align}
		k_\sing &= k_\mathrm{CR}\cos^2\theta \\
		k_{\trip_z} &= k_\mathrm{CR}\sin^2\theta \\
		\epsilon_\sing &= 2J + \delta \epsilon \cos^2\theta \\
		\epsilon_{\trip_z} &= \delta \epsilon \sin^2\theta.
	\end{align}
	The coupling term is given by
		\begin{align}
		\pL_\mathrm{c} = -\left\{\frac{k_\mathrm{CR}}{2}\sin(2\theta) \op{\Pi}_{\sing\trip_z},\ \cdot\  \right\} - i [{\delta\epsilon}\sin(2\theta) \op{\Pi}_{\sing\trip_z}, \ \cdot\  ].
	\end{align} 
	We can expand the reaction superoperator to first-order in the coupling term, which gives the following expression for the reaction superoperator
	\begin{align}
		\pK &\approx \pK^{(0)}  + \pK^{(1)} \label{wsoc-K-eq}\\
		\pK^{(0)} &= - k_\mathrm{f}(1 + k_\mathrm{b} \pL_{\mathrm{d}}^{-1}) \\
		\pK^{(1)} &=  k_\mathrm{f} k_\mathrm{b} \pL_{\mathrm{d}}^{-1}\pL_{\mathrm{c}}\pL_{\mathrm{d}}^{-1}.
	\end{align}
	The $\pK^{(0)}$ is given by
	\begin{align}
		\begin{split}
			\pK^{(0)} &= -\left\{\frac{\tilde{k}^{(0)}_\sing}{2}\op{P}_\sing +\frac{\tilde{k}^{(0)}_{\trip_z}}{2}\op{P}_{\trip_z},\ \cdot\  \right\} \\
			&-i\left[ {\tilde{\epsilon}^{(0)}_\sing}\op{P}_\sing +{\tilde{\epsilon}^{(0)}_{\trip_z}}\op{P}_{\trip_z},\ \cdot\  \right]  \\
			&+\gamma_\sing^{(0)} \left(\op{P}_\sing\ \cdot \ \op{P}_\sing - \frac{1}{2}\left\{\op{P}_\sing,\ \cdot \ \right\}\right) \\
			&+\gamma_{\trip_z}^{(0)} \left(\op{P}_{\trip_z}\ \cdot \ \op{P}_{\trip_z} - \frac{1}{2}\left\{\op{P}_{\trip_z},\ \cdot \ \right\}\right)
		\end{split}
	\end{align}
	where the effective rate constants are given by
	\begin{align}
		\tilde{k}_\sing^{(0)} &= \frac{k_\mathrm{f}k_\sing}{k_\sing + k_\mathrm{b}} \\
		\tilde{k}_{\trip_z}^{(0)} &= \frac{k_\mathrm{f}k_{\trip_z}}{k_{\trip_z} + k_\mathrm{b}}
	\end{align}
	the effective energy shifts are given by
	\begin{align}
		\begin{split}
		\tilde{\epsilon}_\sing^{(0)} &= 2 k_\mathrm{f} k_\mathrm{b} \bigg(\frac{\epsilon _\sing-\epsilon _{\trip_z}}{\left(2 k_\mathrm{b}+k_\sing+k_{\trip_z}\right){}^2+4 \left(\epsilon _\sing-\epsilon _{\trip_z}\right){}^2}\\
		&+\frac{\epsilon _\sing}{\left(2 k_\mathrm{b}+k_\sing\right){}^2+4 \epsilon _\sing^2}\bigg) 
		\end{split} \\
		\begin{split}
		\tilde{\epsilon}_{\trip_z}^{(0)} &=2k_\mathrm{f} k_\mathrm{b} \bigg( \frac{\epsilon _{\trip_z}-\epsilon _\sing}{\left(2 k_\mathrm{b}+k_\sing+k_{\trip_z}\right){}^2+4 \left(\epsilon _\sing-\epsilon _{\trip_z}\right){}^2}\\
		&+\frac{\epsilon _{\trip_z}}{\left(2 k_\mathrm{b}+k_{\trip_z}\right){}^2+4 \epsilon _{\trip_z}^2} \bigg)
		\end{split}
	\end{align}
	and the decoherence rates are given by
	\begin{align}
		\begin{split}
		\gamma_{\sing}^{(0)} &= k_\mathrm{f}-\frac{k_\mathrm{f}k_\mathrm{b} \left(2 k_\mathrm{b}+k_{\trip_z }+k_{\sing }\right)}{\left(2 k_\mathrm{b}+k_{\trip_z }+k_{\sing }\right){}^2+4 \left(\epsilon _{\trip_z }-\epsilon _{\sing }\right){}^2}\\
		&-\frac{k_\mathrm{f}k_\mathrm{b} \left(2 k_\mathrm{b}+k_{\sing }\right)}{\left(2 k_\mathrm{b}+k_{\sing }\right){}^2+4 \epsilon _{\sing }^2} 
		\end{split}\\
	\begin{split}
		\gamma_{\trip_z}^{(0)} &= k_\mathrm{f}-\frac{k_\mathrm{f} k_\mathrm{b} \left(2 k_\mathrm{b}+k_{\sing}+k_{\trip_z}\right)}{\left(2 k_\mathrm{b}+k_{\sing}+k_{\trip_z}\right){}^2+4 \left(\epsilon _{\sing}-\epsilon _{\trip_z}\right){}^2}\\
		&-\frac{k_\mathrm{f}k_\mathrm{b} \left(2 k_\mathrm{b}+k_{\trip_z}\right)}{\left(2 k_\mathrm{b}+k_{\trip_z}\right){}^2+4 \epsilon _{\trip_z}^2}.
		\end{split}
	\end{align}

	Ignoring any decoherence corrections to $\pK^{(1)}$, $\pK^{(1)}$ can be written as
	\begin{widetext}
	\begin{align}
		\pK^{(1)} \approx -\left\{\frac{\tilde{k}^{(1)}_z}{2}\Delta \op{S}_z +\frac{\tilde{k}^{(1)}_{\sing \trip_z}}{2}\op{\Pi}_{\sing\trip_z},\ \cdot\  \right\} -i\left[  {\delta\tilde{\epsilon}^{(1)}_{\sing\trip_z}}\op{\Pi}_{\sing\trip_z},\ \cdot\  \right] 
	\end{align}
	where the spin polarisation selective rate is given by
	\begin{align}
		\tilde{k}^{(1)}_z = -\frac{2 \sin(2\theta) k_\mathrm{f} k_\mathrm{b} \left(2 k_\mathrm{b}+k_\sing+k_{\trip_z}\right) \left(\delta \epsilon  (k_\sing- k_{\trip_z}) +k_\mathrm{CR} \left(\epsilon _{\trip_z}-\epsilon _\sing\right) \right)}{\left(k_\mathrm{b}+k_\sing\right) \left(k_\mathrm{b}+k_{\trip_z} \right) \left(\left(2 k_\mathrm{b}+k_\sing+k_{\trip_z}\right){}^2+4 \left(\epsilon _\sing-\epsilon _{\trip_z}\right){}^2\right)}
	\end{align}
	the spin coherence selective rate is given by
	\begin{align}
	\tilde{k}^{(1)}_{\sing \trip_z} = \frac{\sin(2\theta)k_\mathrm{f}k_\mathrm{b}(4 k_\mathrm{CR} k_\mathrm{b} (k_\sing +  k_{\trip_z}+k_\mathrm{b})+k_\mathrm{CR} (k_\sing + k_{\trip_z})^2+4 \delta \epsilon  (k_\sing-k_{\trip_z})(\epsilon_{\sing} - \epsilon_{\trip_z}) )}{\left(k_\mathrm{b}+k_\sing\right) \left(k_\mathrm{b}+k_{\trip_z}\right) \left(\left(2 k_\mathrm{b}+k_\sing+k_{\trip_z}\right){}^2+4 \left(\epsilon _\sing-\epsilon _{\trip_z}\right){}^2\right)}
	\end{align}
	and the Hamiltonian correction term is given by
	\begin{align}\label{dH1-eq}
		\begin{split}
		\delta\tilde{\epsilon}^{(1)}_{\sing\trip_z} &= \frac{2}{3} \sin(2\theta) k_\mathrm{f}  k_\mathrm{b} \bigg(\frac{2 \left(\left(2 k_\mathrm{b}+k_{\trip_z}\right) \left(2 \delta \epsilon  k_\mathrm{b}+\delta \epsilon  k_\sing-k_\mathrm{CR} \epsilon _\sing\right)-\epsilon _{\trip_z} \left(2 k_\mathrm{CR} k_\mathrm{b}+k_\mathrm{CR} k_\sing+4 \delta \epsilon  \epsilon _\sing\right)\right)}{\left(\left(2 k_\mathrm{b}+k_\sing\right){}^2+4 \epsilon _\sing^2\right) \left(\left(2 k_\mathrm{b}+k_{\trip_z}\right){}^2+4 \epsilon _{\trip_z}^2\right)}\\
		&+\frac{4 \delta \epsilon  k_\mathrm{b}^2+4 \delta \epsilon  k_\mathrm{b} \left(k_\sing+k_{\trip_z}\right)+\delta \epsilon  k_\sing^2+k_\sing \left(-k_\mathrm{CR} \epsilon _\sing+2 \delta \epsilon  k_{\trip_z}+k_\mathrm{CR} \epsilon _{\trip_z}\right)+k_{\trip_z} \left(k_\mathrm{CR} \epsilon _\sing+\delta \epsilon  k_{\trip_z}-k_\mathrm{CR} \epsilon _{\trip_z}\right)}{\left(k_\mathrm{b}+k_\sing\right) \left(k_\mathrm{b}+k_{\trip_z}\right) \left(\left(2 k_\mathrm{b}+k_\sing+k_{\trip_z}\right){}^2+4 \left(\epsilon _\sing-\epsilon _{\trip_z}\right){}^2\right)}\bigg).
		\end{split}
	\end{align}
\end{widetext}
	\bibliography{bibliography.bib}
	\end{document}